\documentclass[%
reprint,
superscriptaddress,
preprintnumbers,
amsmath,amssymb,
aps,
]{revtex4-1}

\usepackage{color}
\usepackage{graphicx}
\usepackage{dcolumn}
\usepackage{multirow}
\usepackage{bm}
\usepackage{url}

\begin{document}
\title{Anisotropy of exchange stiffness based on atomic-scale magnetic properties \\ in rare-earth permanent magnet Nd$_2$Fe$_{14}$B}

\author{Yuta~\surname{Toga}}
\affiliation{ESICMM, National Institute for Materials Science (NIMS), Tsukuba, Japan}

\author{Masamichi~\surname{Nishino}}
\affiliation{Research Center for Advanced Measurement and Characterization, National Institute for Materials Science (NIMS), Tsukuba, Japan}
\affiliation{ESICMM, National Institute for Materials Science (NIMS), Tsukuba, Japan}

\author{Seiji~\surname{Miyashita}}
\affiliation{Department of Physics,~The University of Tokyo, Tokyo, Japan}
\affiliation{ESICMM, National Institute for Materials Science (NIMS), Tsukuba, Japan}

\author{Takashi~\surname{Miyake}}
\affiliation{CD-FMat, National Institute of Advanced Industrial Science and Technology (AIST), Tsukuba, Japan}
\affiliation{ESICMM, National Institute for Materials Science (NIMS), Tsukuba, Japan}

\author{Akimasa~\surname{Sakuma}}
\affiliation{Department~of~Applied Physics, Tohoku~University, Sendai, Japan}

\,\
\date{\today}

\begin{abstract}

We examine the anisotropic properties of the exchange stiffness constant, $\mathcal{A}$,
for rare-earth permanent magnet, Nd$_2$Fe$_{14}$B, by connecting analyses with two different scales of length, i.e.,
Monte Carlo (MC) method with an atomistic spin model and Landau-Lifshitz-Gilbert (LLG) equation with a continuous magnetic model.
The atomistic MC simulations are performed on the spin model of Nd$_2$Fe$_{14}$B constructed from ab-initio calculations, 
and the LLG micromagnetics simulations are performed with the parameters obtained by the MC simulations.
We clarify that the amplitude and the thermal property of $\mathcal{A}$ depend on the orientation in the crystal, which are attributed to the layered structure of Nd atoms and weak exchange couplings between Nd and Fe atoms.
We also confirm that the anisotropy of $\mathcal{A}$ significantly affects the threshold field for the magnetization reversal (coercivity) given by the depinning process.

\end{abstract}

\pacs{test}

\maketitle

\section{Introduction}
In rare-earth permanent magnets such as Nd$_2$Fe$_{14}$B, SmCo$_5$, Sm$_2$Fe$_{17}$N$_3$, and so on, 
the rare-earth elements and the transition metals combine in atomic scale to produce 
strong magnetic anisotropy and strong ferromagnetic order.
The strong magnetic anisotropy originated from the $4f$-electrons of the rare-earth atom is maintained at room temperature (RT) by an interaction with the strong ferromagnetic order of the transition metals\,\cite{buschow_intermetallic_1977,miyake_quantum_2018}.
However, in Nd$_2$Fe$_{14}$B which is well known as the highest performance permanent magnet at RT\,\cite{sagawa_permanent_1984,croat_prfe_1984,herbst_relationships_1984}, 
improvement of coercivity at high temperature is required for industrial application\,\cite{hirosawa_perspectives_2017}
because it has a relatively low Curie temperature ($\sim 585\,\rm K$) as compared with the other rare-earth magnets.

Coercivity mechanism of Nd$_2$Fe$_{14}$B and the other rare-earth magnets has not been fully elucidated yet.
To analyze intrinsic magnetic properties of the rare-earth magnets (e.g., magnetization, magnetic anisotropy, exchange stiffness, Curie temperature, etc.),
we should consider the inhomogeneous magnetic structures and thermal fluctuations in an atomic scale ($\sim\rm\AA$).
%
The rare-earth magnets in the practical use are polycrystalline materials composed of the main rare-earth magnet phase ($\sim\rm \mu m$), and magnetic or non-magnetic grain boundary phases ($\sim \rm nm$).
Therefore, the coercivity depends on not only the intrinsic magnetic parameters, such as the anisotropy energy of the main phase but also on the magnetic properties of the grain boundary and microstructure of the main and grain phases\,\cite{givord_physics_2003,fischbacher_micromagnetics_2018}.
%
For this reason, study of the coercivity requires analyses from the atomic scale ($\sim\rm\AA$) to the macroscopic scale ($\sim \mu\rm m$).
This fact has inhibited to elucidate the coercivity in a systematic manner.
%


%
Theoretically analyses for the coercivity in most have been carried out with a continuum model that uses 
the magnetic anisotropy constants, $K_u$, and the exchange stiffness constant, $\mathcal{A}$ as macroscopic parameters (see Eq.~\eqref{eq:hami_cont}).
%
%
In the junction systems consisting of hard and soft magnetic phases\,\cite{paul_general_1982,sakuma_micromagnetic_1990, kronmuller_micromagnetic_2002, mohakud_temperature_2016}, it has been indicated that both $K_u$ and $\mathcal{A}$ have a large effect on incoherent magnetization reversals (i.e., nucleation and depinning) which
reduce the coercivity from the value of the uniform reversal.
In order to elucidate the coercivity mechanism and improve the coercivity at high temperatures,
it is important to clarify the temperature dependences of $K_u$ and $\mathcal{A}$.
Therefore, many researchers in both experimental and theoretical have studied the thermal properties of $K_u$\,\cite{hirosawa_single_1985,yamada_magnetocrystalline_1986, durst_determination_1986,  skomski_finite-temperature_1998,sasaki_theoretical_2015, toga_monte_2016}.
However, regarding $\mathcal{A}$ for Nd$_2$Fe$_{14}$B, experimentally, the values are observed only at several temperatures\,\cite{mayer_inelastic_1991,mayer_inelastic_1992,bick_exchange-stiffness_2013, ono_observation_2014}, and also there is no theoretical estimation of the temperature dependence as far as we know.

The value of $\mathcal{A}$ is given as a macroscopic properties of exchange stiffness of continuous magnets at each temperature.
{This quantity is related to} the domain wall (DW) width\,\cite{chikazumi_physics_1997} and the critical (magnetization reversal) nucleus size\,\cite{givord_physics_2003}.
For some other magnetic materials, YCo$_5$ and $L1_0$-type magnets (CoPt, FePd, FePt), Belashchenko indicated using ab-initio calculation that $\mathcal{A}$ depends on the orientation in the crystals\,\cite{belashchenko_anisotropy_2004}.
And Fukazawa et al. also pointed out the anisotropic $\mathcal{A}$ for Sm(Fe, Co)$_{12}$ compounds\,\cite{fukazawa_first-principles_2018}.
Recently, Nishino et al. examined the temperature dependence of DW width of Nd$_2$Fe$_{14}$B using an atomistic spin model constructed from ab-initio calculations, and indicated that it also has an orientation dependence\,\cite{nishino_atomistic-model_2017}.
%
Although, in Ref.~\cite{belashchenko_anisotropy_2004}, the effect of anisotropic $\mathcal{A}$ on coercivity is also discussed in a phenomenological way,
microscopic understanding of the temperature and orientation dependences of $\mathcal{A}$ is essential
for the coercivity mechanism in Nd$_2$Fe$_{14}$B magnet.

In the present paper,
by a comparison of the results obtained by a continuum model and those obtained by the atomistic spin model constructed from ab-initio calculations (same spin model as Ref.~\cite{toga_monte_2016, nishino_atomistic-model_2017, hinokihara_exploration_2018}),
we study the temperature and the orientation dependence of $\mathcal{A}$ for Nd$_2$Fe$_{14}$B magnet.
The comparison of the two different scale models {is done by making use of} magnetic DW energy.
This scheme has been used successful for FePt \,\cite{hinzke_orientation_2007,hinzke_domain_2008, atxitia_multiscale_2010} and Co \,\cite{moreno_temperature-dependent_2016} magnetic materials.
We find that $\mathcal{A}$ has the anisotropic property in Nd$_2$Fe$_{14}$B.
Moreover, the reason for the anisotropy is attributed to the weak exchange couplings of Nd and Fe atoms.
Additionally, we performed micromagnetics simulations on the configuration in which a soft magnetic phase is attached to (001) plane or (100) plane of a grain of Nd$_2$Fe$_{14}$B.
The simulations predict that the anisotropy of $\mathcal{A}$ reduces 
the coercivity for the former configuration because $\mathcal{A}$ of $z$-direction is larger than that of $x$-direction,
while it enhances the coercivity for the latter configuration.
The series of the calculation schemes in the present study corresponds to the multiscale analysis\,\cite{miyashita_perspectives_2018, westmoreland_multiscale_2018} that connects different scales from ab-initio calculation to coercivity as macroscopic physics.



%




\section{Models and Method}

\subsection{Atomistic Spin Model}

To include the information of electronic states at an atomistic scale,
we treat the (spin) magnetic moment of each atom as a classical spin and then construct a classical Heisenberg model.
The atomistic Hamiltonian has the form:
%
\begin{eqnarray}
 {\cal H} &=& -2\sum_{i<j} \tilde{J}_{ij} {\bm{e}}_i \cdot {\bm{e}}_j\nonumber \\
   &&-\sum_{i \in {\rm Fe} } D_{i} (e^z_i)^2
   +\sum_{i \in {\rm Nd}} \sum_{l=2,4,6} B_{l,i}^{m_l} \hat{\mathcal{O}}^{m_l}_{l,i},
   \label{eq:hami}
\end{eqnarray}
where
$\tilde{J}_{ij}$ is the Heisenberg exchange coupling constants including the spin amplitudes ($S_iS_j$) between the $i$th and $j$th sites,
and ${\bm e}_i$ is the normalized spin moment at the $i$th site.
The coefficient, $D_i$, in the second term denotes the strength of the magnetic anisotropy of Fe sites.
The third term is the magnetic anisotropy of Nd sites, which is formulated by the crystal field theory for $4f$-electrons\,\cite{stevens_matrix_1952,yamada_crystal-field_1988},
$B_{l,i}^{m_l}$ is the crystal electric field coefficient and $\hat{\mathcal{O}}^{m_l}_{l,i}$ is the Stevens operator.
%
In the present study, $B_{l,i}^{m_l}$ takes a fixed value,
whereas $\hat{\mathcal{O}}_{l,i}^{m_l}$ depends on the state of a total angular momentum of $4f$-electrons, $\bm{\mathcal{J}}_i$.
Here, we fix $\bm{\mathcal{J}}_i$ parallel to the normalized spin moment on the $i$th site,
i.e. $\bm{\mathcal{J}}_i = \mathcal{J}_i\bm{e}_i$ ($\mathcal{J}_i=9/2$ for Nd atom).
For simplicity, we consider only the diagonal terms $m_l=0$.

\begin{figure}[t]
\includegraphics[width=8.6cm]{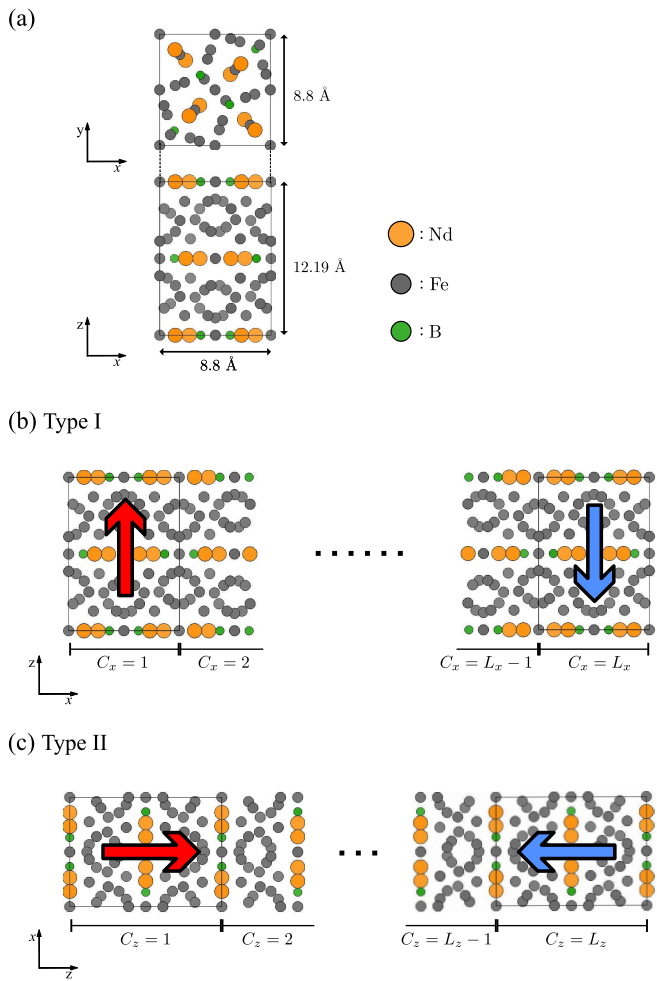}
\caption{\label{fig:str}
(a) Unit Cell of Nd$_2$Fe$_{14}$B including 68 atoms\,\cite{herbst_relationships_1984}.
Two types of spin configurations for Nd$_2$Fe$_{14}$B atomistic spin model: DW orientation is along (b) the $yz$-plane (Type\,I) and (c) the $xy$-plane (Type\,II).
The arrows in figures describe the antiparallel (AP) boundary condition.
These crystal structures were plotted using {\sc vesta}\,\cite{momma_vesta_2011}.
}
\end{figure}

For these input parameters of Nd$_2$Fe$_{14}$B magnet, we adopt the same values in the previous studies\,\cite{toga_monte_2016, nishino_atomistic-model_2017,hinokihara_exploration_2018}.
Exchange coupling constants, $\tilde{J}_{ij}$, were calculated with Liechtenstein's formula\,\cite{liechtenstein_local_1987} on the Korringa-Kohn-Rostoker (KKR) Green's-function code, {\sc machikaneyama (akaikkr)}\,\cite{akai_akaikkrmachikaneyama_nodate}.
In the present study, to reduce computational cost, we use only short-range exchange couplings within the range of $r_{\rm cut}=3.52\,\rm\AA$, in which primary Fe--Fe and Nd--Fe interactions are included.
Anisotropy terms, $D_{i}$ and $B_{l,i}^{m_l}$, were determined from the previous first-principles calculation\,\cite{miura_magnetocrystalline_2014} and the experimental result\,\cite{yamada_crystal-field_1988}, respectively.
Consequently, the atomistic spin model uses many input parameters in a unit cell which includes 68 atoms (see Fig.~\ref{fig:str}\,(a)).
The previous study\,\cite{toga_monte_2016} confirmed that the model and parameters are highly reliable for the magnetic properties of the Nd$_2$Fe$_{14}$B magnet.





\subsection{Continuum Model}

%
Under the continuum approximation, the micromagnetic energy of the exchange couplings and the magnetic anisotropies at temperature $T$ is expressed as follows\,\cite{miltat_numerical_2007}:
\begin{eqnarray}
E_{\rm cont}&=& 
\int_{\rm V} d^3\bm{r}\,\left[ \sum_{l=x,y,z} \mathcal{A}_l(T) \left(\nabla_l {\bm m}({\bm r})\right)^2 \right] \nonumber \\
  && + \int_{\rm V} d^3\bm{r}\, \mathcal{E}_{K}(T,\theta({\bm r})),
  \label{eq:hami_cont}
\end{eqnarray}
where ${\bm m}$ denotes a normalized magnetization vector, $\mathcal{A}_l(T)$ is the exchange stiffness constant of each direction ($x,y,z$) in the crystal,
and $\mathcal{E}_{K}$ is the magnetic anisotropy energy which is usually expressed as:
\begin{eqnarray}
\mathcal{E}_{K}(T,\theta) &=& K_1(T) \sin^2{\theta}+K_2(T) \sin^4{\theta}+K_4(T) \sin^6 {\theta}, \nonumber\\
   \label{eq:hami_ea}
\end{eqnarray}
where $K_u(T)\ (u=1, 2, 4)$ are the magnetic anisotropy constants and $\theta$ is the angle of magnetization measured from $z$-axis.
The continuum model uses the temperature-dependent parameters to express the thermal effects instead of thermal fluctuations in the atomistic model.

Micromagnetic simulations\,\cite{nakatani_direct_1989,miltat_numerical_2007} have been carried out based on the continuum approximation and Landau-Lifshitz-Gilbert (LLG) equation which describes time evolution of magnetization relaxation process (see Sec.\ref{sec:llg}).
In the practical simulation, the real space in Eq.~\eqref{eq:hami_cont} is discretized with a grid whose width should be smaller than a DW width and a magnetostatic exchange length\,\cite{miltat_numerical_2007}.
%
In most micromagnetic simulations for Nd$_2$Fe$_{14}$B magnet, each grid size was set to ($1{\mathchar`-}2\, \rm nm)^3$, and as the input parameters,
experimentally observed values of $\mathcal{A}_l(T)$ and $K_u(T)$ were used.

In the continuum model,
many input parameters of Eq.~\eqref{eq:hami} in atomic scale are expected to be renormalized in the few macroscopic parameters of Eq.~\eqref{eq:hami_cont} at each temperature.


 
\subsection{Methods for determining $K_u(T)$ and $\mathcal{A}(T)$}

In the present study,
the macroscopic parameters are evaluated by comparing the energies of the DW and of the magnetic anisotropy obtained by the above two models at each temperature.
We use a Monte Carlo (MC) scheme to calculate Helmholtz free energies on the atomistic spin model.
%
%
%
In order to evaluate $\mathcal{A}$, we consider the two quantities obtained from DW energy:
$\mathcal{E}_{\rm dw}(T)$ and $\mathcal{F}_{\rm dw}(T)$\,\cite{chikazumi_physics_1997,hinzke_domain_2008}.
$\mathcal{E}_{\rm dw}$ is the DW energy in the continuum model, which is
expressed by the following formula:
\begin{eqnarray}
 \mathcal{E}_{\rm dw}(T)&=&2\sqrt{\mathcal{A}(T)} \int_0^\pi d\theta\,\sqrt{\mathcal{E}_{K}(T,\theta)}.
 \label{eq:dwene_rel}
\end{eqnarray}
To obtain $\mathcal{A}(T)$, we need the values of $\mathcal{E}_{\rm dw}(T)$ and $\mathcal{E}_{K}(T,\theta)$. 
We regard $\mathcal{E}_{\rm dw}(T)$ to be equal to 
the DW free energy in the atomistic spin model, $\mathcal{F}_{\rm dw}(T)$,
which is expressed by the following formula:
\begin{eqnarray}
 \mathcal{F}_{\rm dw}(T)&=& T\int_T^{\infty} dT^\prime\, \frac{E_{\rm dw}(T^\prime)}{(T^\prime)^2}.
 \label{eq:dwene_cal}
\end{eqnarray}
%
The DW internal energy, $ E_{\rm dw}(T)$, is defined as an energy difference between the internal energies obtained in different boundary conditions with and without DW.
For this calculation, we fix the boundaries in the direction antiparallel (AP) or parallel (PA).
Additionally, since the crystal structure of Nd$_2$Fe$_{14}$B has anisotropy,
we consider two models with the two fixed boundaries depicted in Fig.~\ref{fig:str}(b) and (c).
Thus, we have two types of DWs\,\cite{nishino_atomistic-model_2017}.
One is a DW along the $yz$-plane (type\,I), and the other is along the $xy$-plane (type\,II).
We set the periodic boundary condition in the $yz (xy)$-plane for the model of type I (type II).

%
%
%

To calculate $\mathcal{E}_K(T,\theta)$, we adopt the constrained MC (C-MC) method for the atomistic spin model.
The C-MC method samples spin configurations under the condition of a fixed angle of total magnetization,
which allows us to calculate the angle dependence of the spin torque and the free energy\,\cite{asselin_constrained_2010}.
%
By considering that $\mathcal{E}_K$ is equal to the free energy of magnetic anisotropy, we calculate $\mathcal{E}_K$ from the atomistic model.
The previous study\,\cite{toga_monte_2016} showed the validity of the C-MC methods for calculating $K_u(T)$ in $\mathcal{E}_K(T,\theta)$ of the Nd$_2$Fe$_{14}$B atomistic spin model. 
Note that if Eq.~$(\ref{eq:hami_ea})$ is given in the form $\mathcal{E}_K=K_1 \sin^2 \theta$ ($K_2=K_4=0$),
Eq.~$(\ref{eq:dwene_rel})$ can be integrated analytically and gives the well known relation: $\mathcal{E}_{\rm dw}(T)=4\sqrt{\mathcal{A}(T) K_1(T)}$.


In the present paper, the MC and C-MC simulations run 100000-200000 MC steps for equilibrium and the following 100000-200000 MC steps for taking statistical averages,
where a MC step corresponds to one trial for each spin to be updated.
We calculated the average of magnetic anisotropy energies and DW energies from 8-20 different runs with different initial conditions and random sequences.

%



\section{Results and Discussion}

\subsection{Magnetic anisotropy and domain wall energy}

\begin{figure}[t]
\includegraphics[width=8.6cm]{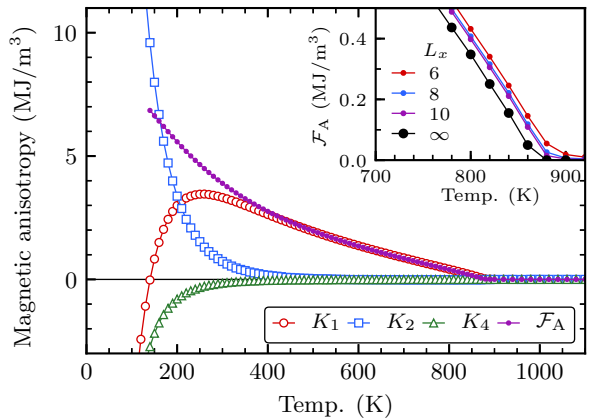}
\caption{\label{fig:ani}
Magnetic anisotropy constants and $\mathcal{F}_{\rm A}$ as a function of temperature for the system of $10\times10\times 7$ unit cells.
Inset show $\mathcal{F}_{\rm A}(T)$ for each system size, $(L_x,L_y,L_z)=(6,6,4)$, $(8,8,6)$, $(10,10,7)$, and the extrapolation data ($L_x \propto N^{1/3} \rightarrow \infty$).
}
\end{figure}

As mentioned above, the evaluation of the exchange stiffness constant requires precise calculations of the magnetic anisotropy constants and the DW energy.
%
First, we show the temperature dependence of the magnetic anisotropy constants calculated by the C-MC method in Fig.~\ref{fig:ani}, which qualitatively agrees with previous experiments \cite{yamada_magnetocrystalline_1986,durst_determination_1986}.
There, we used the system of $10\times10\times 7$ unit cells imposing the periodic boundary conditions.
In uniaxial anisotropy, we study the quantities:
$\mathcal{F}_{\rm A}(T) \equiv \mathcal{E}_{K}(T,\theta=\pi/2) = \mathcal{E}_K(T,\pi/2) - \mathcal{E}_K(T,0) = \sum_u K_u(T)$.
Figure~\ref{fig:ani} shows that $\mathcal{F}_{\rm A}(T)$ rapidly decreases with the temperature below $400\,\rm K$.
This temperature dependence is understood as a result of fragile thermal properties of Nd atoms.
That is, while at low temperature,
%
%
the anisotropy of Nd atoms is dominant,
the anisotropy of Nd atoms decreases with the temperature more rapidly than that of Fe atoms,
because the exchange field of a Nd atom (sum of exchange couplings that connect to a Nd atom) is much smaller (about $20\mathchar`-25\,\%$) than that of a Fe atom
(see also the detailed discussion in Ref.~\cite{toga_monte_2016}).



Additionally, owing to the higher order terms of the Nd anisotropy, $B_4^0, B_6^0$, the magnetization of Nd$_2$Fe$_{14}$B is tilted from the $z$-axis at low temperatures.
This fact causes that the deviation from uniaxial anisotropy occurs in the region of $-2K_2<K_1 < 0$\,\cite{smit_ferrites_1959}.
%
%
Note that Eq.~\eqref{eq:dwene_rel} is defined under the condition of uniaxial anisotropy.
And thus, in the present paper, we discuss the DW energy and the exchange stiffness in the region of $T\geq 200\, \rm K$.

\begin{figure}[t]
\includegraphics[width=8.6cm]{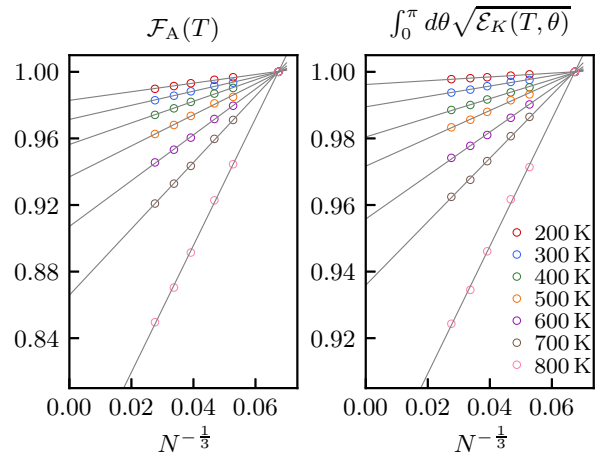}
\caption{\label{fig:extra}
Finite size extrapolation of $\mathcal{F}_{\rm A}(T)$ and $\int_0^\pi d\theta \sqrt{\mathcal{E}_K(T,\theta) }$. Solid lines are linear fits to the Monte Carlo results for each system size ($N$ is the total number of spins). Vertical axes are normalized at the values for $N=3264$.
}
\end{figure}

As a matter of practice, finite-size effect of numerical results about the magnetic anisotropy is significant for the evaluation of the exchange stiffness and the DW width, especially at high temperatures.
To avoid this problem, we extrapolate $\mathcal{F}_{\rm A}(T)$ and $\int_0^\pi d\theta \sqrt{\mathcal{E}_K(T,\theta) }$ [these are used in Eqs.~\eqref{eq:dww} and \eqref{eq:dwene_rel}] to the thermodynamic limit ($N \rightarrow \infty$)
using linear functions in $N^{-1/3}$.
Figure~\ref{fig:extra} shows the typical fitting results which indicate that the MC results are well fitted with the linear functions.
The extrapolated results for $\mathcal{F}_{\rm A}$ near the Curie temperature, $T_{\rm C}$, are summarized in the inset of Fig.~\ref{fig:ani}, where we fix as $K_u=0$ above $T_{\rm C}\, (=870\,\rm K)$.
We use these extrapolated results to evaluate the exchange stiffness and the DW width.


\begin{figure}[t]
%
(a) Type\,I ($yz$-plane)\\
\includegraphics[width=8.6cm]{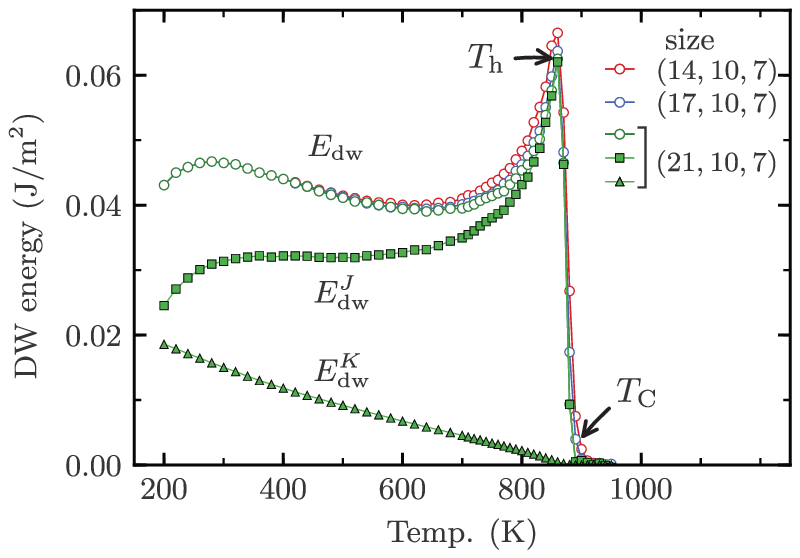}\\
  
(b) Type\,II ($xy$-plane)\\
\includegraphics[width=8.6cm]{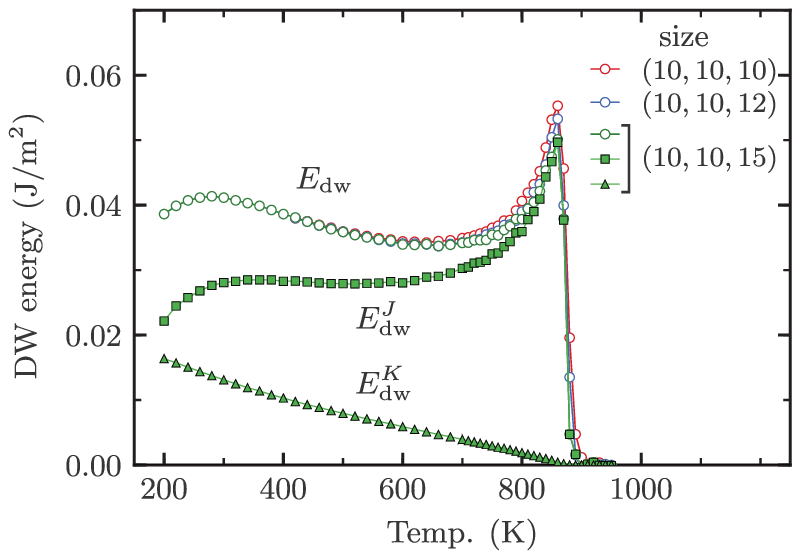}\\
\caption{
Temperature dependence of DW internal energy for three system sizes, $(L_x,L_y,L_z)$, in the DWs of (a) Type I and (b) Type II.
Total DW energy is divided into anisotropy and exchange terms, $E_{\rm dw}=E_{\rm dw}^K+E_{\rm dw}^J$, for the largest system size of each DW type.
\label{fig:diene}
}
\end{figure}

Next, we focus on the DW energy.
Figure~\ref{fig:diene} shows the temperature dependence of the DW internal energy, $E_{\rm dw}$, for the two DW types (see Fig.~\ref{fig:str}).
We also plot $E_{\rm dw}$ for three different system sizes in the directions perpendicular to the DWs, i.e. $L_{x}=$14, 17, and 21 $(L_{z}=$10, 12, and 15), for type I (type II).
From these results, it is confirmed that these systems are large enough to calculate $E_{\rm dw}$.
To analyze in detail the temperature dependence, we divided $E_{\rm dw}$ into the magnetic anisotropy term ($E_{\rm dw}^K$) and the exchange term ($E_{\rm dw}^J$), and plot the contributions in Fig.~\ref{fig:diene}.
The DW energies of type\,I and type\,II show a qualitatively similar temperature dependence.
Both energy terms naturally vanish for $T \geq T_{\rm C}$, because DW does not appear even in the AP boundary condition.
For $T < T_{\rm C}$,
the anisotropy term decreases monotonically as the temperature increases whereas the exchange term increases and takes a peak value at the temperature, $T_{\rm h}$, slightly below $T_{\rm C}$.

The temperature dependences are interpreted as follows.
The monotonic decrease of $E_{\rm dw}^K$ is merely due to the decrease of the thermally averaged magnetic moments,
which also corresponds to the decrease of $\mathcal{F}_{\rm A}(T)$ in Fig.~\ref{fig:ani}.
The magnetic anisotropy energy depends on the angle from $z$-axis of each spin,
whereas the exchange coupling energy depends on the relative angle of spin pairs.
%
%
%
Now, the DW energy is defined as the difference of the internal energies between the PA ($E_{\rm PA}$) 
and the AP ($E_{\rm AP}$) boundary conditions,
i.e., $E_{\rm dw}(T)=E_{\rm PA}-E_{\rm AP}$.
$E_{\rm dw}^J$ is the part coming from the exchange term,
thus it is influenced by the difference between the two boundary conditions in the fluctuations of the relative angles of spin pairs.
In the configuration of the AP boundary condition, DW exists.
In the DW, the ferromagnetic order is weakened because the spin configuration is forcibly twisted,
and the DW has a spiral non-collinear structure with the perpendicular ($xy$) component at low temperatures.
This fact gives the difference of the energies $E_{\rm dw}^J$.
We expect that $E_{\rm dw}^J$ due to the formation of the non-collinear structure would be enhanced near the critical point where the width of DW increases.

%

At a temperature $T_{\rm h}$ which is slightly lower than $T_{\rm C}$,
the profile of perpendicular ($xy$) moments is destroyed by the thermal fluctuation.
As a thermally averaged magnetic structure, the non-collinear structure (the spiral structure) becomes a collinear structure.
At this point, $E_{\rm dw}^J$ takes the peak value at $T_{\rm h}$ and then decreases rapidly and approaches zero towards $T_{\rm C}$.
This break of the non-collinear structure was discussed in previous studies as the disappearance of an $xy$--magnetization inside the DW\,\cite{bulaevkivi_temperature_1964,kazantseva_transition_2005,hinzke_domain_2008}.
%

%



\begin{figure}[t]

\includegraphics[width=8.6cm]{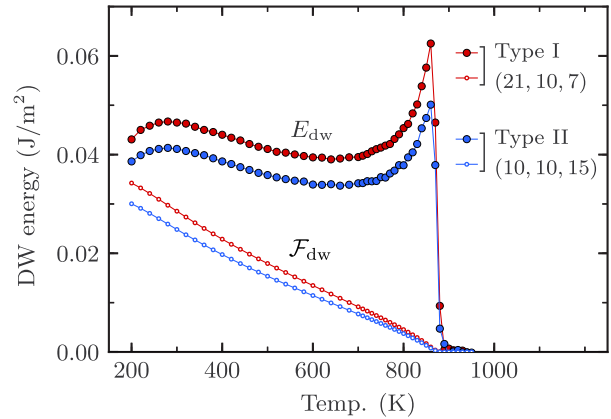}
\caption{\label{fig:dfene}
Temperature dependence of DW free energy, $\mathcal{F}_{\rm dw}$ and internal energy, $E_{\rm dw}$ (same as Fig.~\ref{fig:diene} (a) and (b)), for each DW type.
}
\end{figure}

By those internal DW energies and Eq.~\eqref{eq:dwene_cal}, we can calculate DW free energies, $\mathcal{F}_{\rm dw}$.
In Fig.~\ref{fig:dfene}, we plot the temperature dependence of $\mathcal{F}_{\rm dw}$ and $E_{\rm dw}$ (the same data shown in Fig.~\ref{fig:diene}) for both DW types.
The differences between $\mathcal{F}_{\rm dw}$ and $E_{\rm dw}$ correspond to the contribution of magnetic entropy.
%
The difference in the DW energy between type I and type II naturally indicates an anisotropy concerning to the direction in which DWs are generated.
The DW prefers to be generated in the configuration of type II.
These observations imply the magnetization reversal starts from the $z$-plane.
Moreover, the generation of the DW in the magnets would also depend on the grain boundary phase and the dipole-dipole interaction.
We will discuss these properties in more realistic magnetization reversal process in Sec.~\ref{sec:llg}.



\subsection{Exchange stiffness constant}

\begin{figure}[t]
\includegraphics[width=8.6cm]{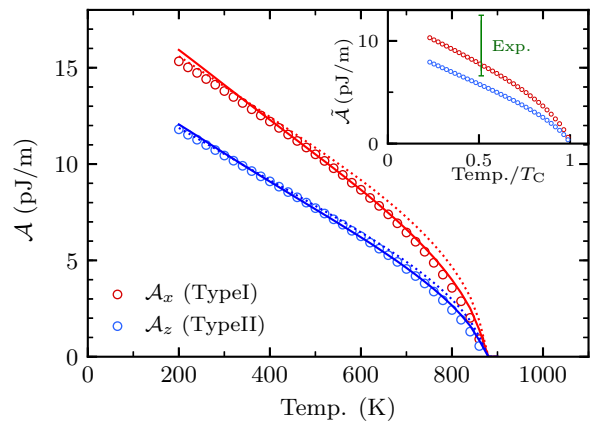}
\caption{\label{fig:stiffness}
Temperature dependence of exchange stiffness constant, $\mathcal{A}$ for each DW type.
Solid (dotted) lines are fitting results in the range of $200\mathchar`-860\,\rm K$ ($200\mathchar`-400\,\rm K$) using $\mathcal{A}(T)=C M(T)^n$.
Inset show the renormalized values, $\tilde{\mathcal{A}}$ (see Eq.~\ref{eq:aex_renormalize}), and the green bar denotes the range of the experimental values at room temperature\,\cite{bick_exchange-stiffness_2013,ono_observation_2014}.
}
\end{figure}

The exchange stiffness constants, $\mathcal{A}$, for the two directions can be evaluated by using Eq.~\eqref{eq:dwene_rel} and the numerical results for the magnetic anisotropy and the DW energy,
whose temperature dependence are shown in Fig.~\ref{fig:stiffness}.
Here, we define the value of $\mathcal{A}$ calculated from the configuration of type I (type II) as the exchange stiffness constant of the $x\, (z)$-direction, $\mathcal{A}_{x (z)}$.
Reflecting the anisotropy of the DW energy, the exchange stiffness constant naturally has the anisotropy depending on the direction in the crystal.
For the comparison of $\mathcal{A}$ with experimental values,
it is reasonable to normalize the temperature dependence with $T_{\rm C}$,
because the spin model overestimates ($T^{\rm MC}_{\rm C}\, = 870\,\rm K)$ compared to experiment ($T^{\rm EXP}_{\rm C} \simeq 585\,\rm K$).
In addition, as pointed in the mean field approximation\,\cite{kronmuller_handbook_2007}, $\mathcal{A}$ is roughly proportional to $T_{\rm C}$, so that we also rescale the values of $\mathcal{A}$:
\begin{eqnarray}
  \tilde{\mathcal{A}}(T) = \mathcal{A}(T)
{T^{\rm EXP}_{\rm C}\over T^{\rm MC}_{\rm C}}.
  \label{eq:aex_renormalize}
\end{eqnarray}
The inset of Fig.~\ref{fig:stiffness} shows the rescaled data and experimental values (green bar) at RT\,\cite{bick_exchange-stiffness_2013,ono_observation_2014}.
Although the experimental values have some variation ($6.6\mathchar`-12.5\,\rm pJ/m$), the calculation results are well consistent with the lower experimental values at RT.
 
\begin{figure}[t]
\includegraphics[width=8.6cm]{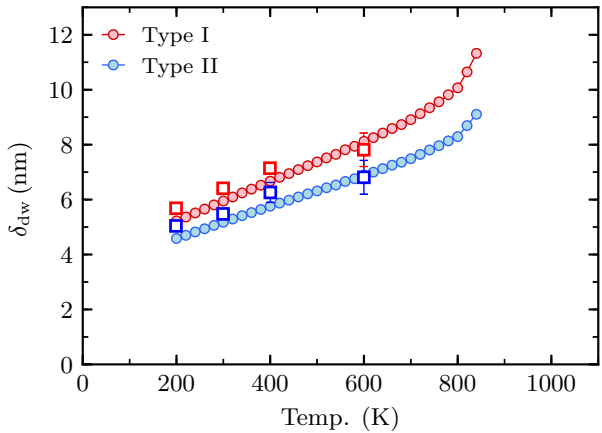}
\caption{\label{fig:dww}
Temperature dependence of DW width, $\delta_{\rm dw}$, for each DW type.
Circle lines are our calculation results from $K_u(T)$ and $\mathcal{F}_{\rm dw}$.
Square points are the previous numerical results which are evaluated directly from the snapshots of spin configurations while in MC simulations.
}
\end{figure}
%
With $\mathcal{A}$ and $K_u$ which have been obtained, the DW width, $\delta_{\rm dw}$, is calculated from the following relation\,\cite{durst_determination_1986, chikazumi_physics_1997} :
\begin{eqnarray}
 \label{eq:dww}
 \delta^{x (z)}_{\rm dw}(T)&=&\pi \left.\frac{\partial \omega(T)}{\partial \theta}\right|_{\theta=\frac{\pi}{2}} 
 =\pi\sqrt{ \frac{\mathcal{A}_{x (z)}(T)}{\mathcal{F}_{\rm A}(T)}}, \\
 \omega(T)&=&\sqrt{\mathcal{A}_{x (z)}(T)}\int_0^\theta \frac{d\theta}{\sqrt{\mathcal{E}_K(T,\theta)}}.\nonumber
\end{eqnarray}
To confirm the validity of our evaluation procedure for $\mathcal{A}$,
in Fig.~\ref{fig:dww}, we compared our results (circle) with those of the previous study (square) \,\cite{nishino_atomistic-model_2017}.

In the previous study, $\delta_{\rm dw}$ was evaluated directly from the snapshots of spin configurations using the same atomistic Hamiltonian of the present study.
Note that, we set the periodic boundary condition in the $yz (xy)$-plane for the model of type I (type II),
whereas the previous calculation was performed under the open-boundary conditions for both models.
%
%
%
Our results of $\delta_{\rm dw}$ qualitatively agree with the previous results,
although they tend to take a smaller value because the thermal fluctuation becomes smaller than the previous study due to the difference of boundary conditions.
The comparisons with the previous experiments and the numerical study as mentioned above guarantee our results concerning to $\mathcal{A}$.

Now, we study its thermal properties.
The temperature dependence of $\mathcal{A}$ is often discussed in relation to magnetization by using the following expression:
\begin{eqnarray}
  \mathcal{A}(T) = \mathcal{A}(0) \left[ \frac{M(T)}{M(0)} \right]^n,
  \label{eq:aex_relation_m}
\end{eqnarray}
where $M(T)$ is the amplitude of the magnetization (not shown, see Ref.~\,\cite{toga_monte_2016}). 
Under a mean field approximation with homogeneous spin systems, the exponent $n$ is 2\,\cite{kronmuller_handbook_2007, atxitia_multiscale_2010}.
%
%
However, 
under more accurate methods,
the exponent takes a value different from $2$, for example, FePt: $n=1.76$ \,\cite{atxitia_multiscale_2010}, hcp-Co: $n=1.79$ \,\cite{moreno_temperature-dependent_2016}.
Thus we estimate $n$ to examine the thermal properties for the Nd$_2$Fe$_{14}$B,
by fitting $\mathcal{A}(T)$ in Fig.~\ref{fig:stiffness} with the form $C M(T)^n$,
where $C$ is a fitting constant.
%
Note that $n$ depends on the fitting range of temperature 
$n_{x(z)}=1.68\,(1.84)$ in the range of $200\mathchar`-860\,\rm K$, whereas $n_{x(z)}=1.46\,(1.69)$ in the range of $200\mathchar`-400\,\rm K$.
Fitting lines of the former and the latter are plotted by the solid and dotted lines in Fig.~\ref{fig:stiffness}, respectively.
Microscopically, Fe and Nd show different thermal properties. Indeed, Nd atoms have weak exchange coupling and so do not have much influence on $\mathcal{A}$,
whereas they have a large magnetic moment ($\sim 2.87\,\mu_B/\rm atom$).
The magnetization of Nd atoms is decreased more rapidly with temperature than that of Fe atoms,
which largely affects the temperature dependence of total magnetization.
Thus, intrinsically $n$ depend on the fitting range largely.
%
However, the important point here is that $n_z$ always takes larger values than $n_x$ regardless of the fitting range (we checked it).
%
This relation implies the macroscopic exchange coupling in the $z$-direction is weaker than that in $x$-direction, not only for the coupling strength but also for thermal tolerance.


%

\begin{figure}[t]
\includegraphics[width=8.6cm]{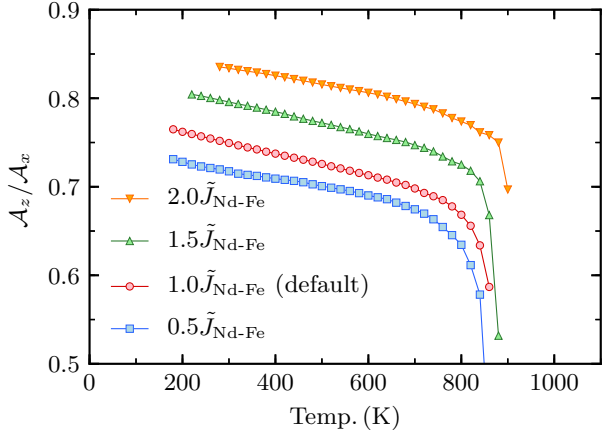}

\caption{\label{fig:ratio}
The anisotropy ratio of the exchange stiffness, $\mathcal{A}_z/\mathcal{A}_x$, as a function of temperature for the four models with different exchange coupling between Nd and Fe, $ \tilde{J}_{\rm Nd \mathchar`- Fe}$.
}
\end{figure}

As the reason for the anisotropy of $\mathcal{A}$ to crystal orientation, 
the crystal structure of Nd$_2$Fe$_{14}$B is naturally invoked.
Nd$_2$Fe$_{14}$B has the layered structure of Fe-layer and NdFe-layer (B has little effect on magnetic properties) along the $z$-axis as shown in Fig.~\ref{fig:str}\,(a).
In Nd$_2$Fe$_{14}$B, exchange couplings ($\tilde{J}_{ij}$ in Eq.~\eqref{eq:hami}) are mainly contributed by bonding between Fe and Fe atoms, $\tilde{J}_{\rm Fe\mathchar`- Fe}$, and between Nd and Fe atoms, $\tilde{J}_{\rm Nd\mathchar`- Fe}$ ($|\tilde{J}_{\rm Nd\mathchar`- Nd}|$ is negligibly small).
Each bond of $\tilde{J}_{\rm Fe\mathchar`- Fe}$ has much larger amplitudes than $\tilde{J}_{\rm Nd\mathchar`- Fe}$
($\tilde{J}_{\rm Fe\mathchar`- Fe}:-4.35 \,\mathchar`-\, 22.34\,\rm meV$, $\tilde{J}_{\rm Nd\mathchar`- Fe}:-0.16 \,\mathchar`-\, 3.55\,\rm meV$).
%
%
Therefore, it is anticipated that the anisotropy of $\mathcal{A}$ comes from the inhomogeneous distribution of the exchange couplings and the atom positions in the crystal structure.
To support this anticipation in detail, we examine the relation between the anisotropy and the strength of $\tilde{J}_{\rm Nd\mathchar`- Fe}$.


Figure~\ref{fig:ratio} shows the ratio of $\mathcal{A}_z$ to $\mathcal{A}_x$ for four models with different values of $\tilde{J}_{\rm Nd \mathchar`- Fe}$.
Beside the default model ($1.0 \tilde{J}_{\rm Nd \mathchar`- Fe}$, the ratio is calculated from Fig.~\ref{fig:stiffness}),
we also calculate other three cases with all the bonds $\tilde{J}_{\rm Nd \mathchar`- Fe}$ reduced by half ($0.5 \tilde{J}_{\rm Nd \mathchar`- Fe}$), increased by half ($1.5 \tilde{J}_{\rm Nd \mathchar`- Fe}$), and doubled ($2.0 \tilde{J}_{\rm Nd \mathchar`- Fe}$).
It is clearly found that $\mathcal{A}$ gets close to isotropic (i.e., $\mathcal{A}_z=\mathcal{A}_x$) as $\tilde{J}_{\rm Nd \mathchar`- Fe}$ increase in the whole temperature range.

As another feature, the ratio slowly decreases with the temperature for all the cases, which
indicates that the temperature dependence of $\mathcal{A}_x$ and $\mathcal{A}_z$ are different.
This difference corresponds to the difference between $n_x$ and $n_z$ in Eq.~\eqref{eq:aex_relation_m}.
As the temperature increases, the contribution of $\tilde{J}_{\rm Nd\mathchar`-Fe}$ to $\mathcal{A}$ becomes smaller than that of $\tilde{J}_{\rm Fe\mathchar`-Fe}$ because the spin moments of Nd atoms are more easily broken by thermal fluctuations compared with those of Fe atoms.
From the above analysis, we conclude that the reason of the anisotropy of $\mathcal{A}$ in Nd$_2$Fe$_{14}$B comes from the weakness of $\tilde{J}_{\rm Nd\mathchar`-Fe}$ and the layered structure of Nd atoms.


\subsection{Effect of anisotropic exchange on coercivity}\label{sec:llg}

\begin{figure}[t]
\vspace{5mm}
\includegraphics[width=8.6cm]{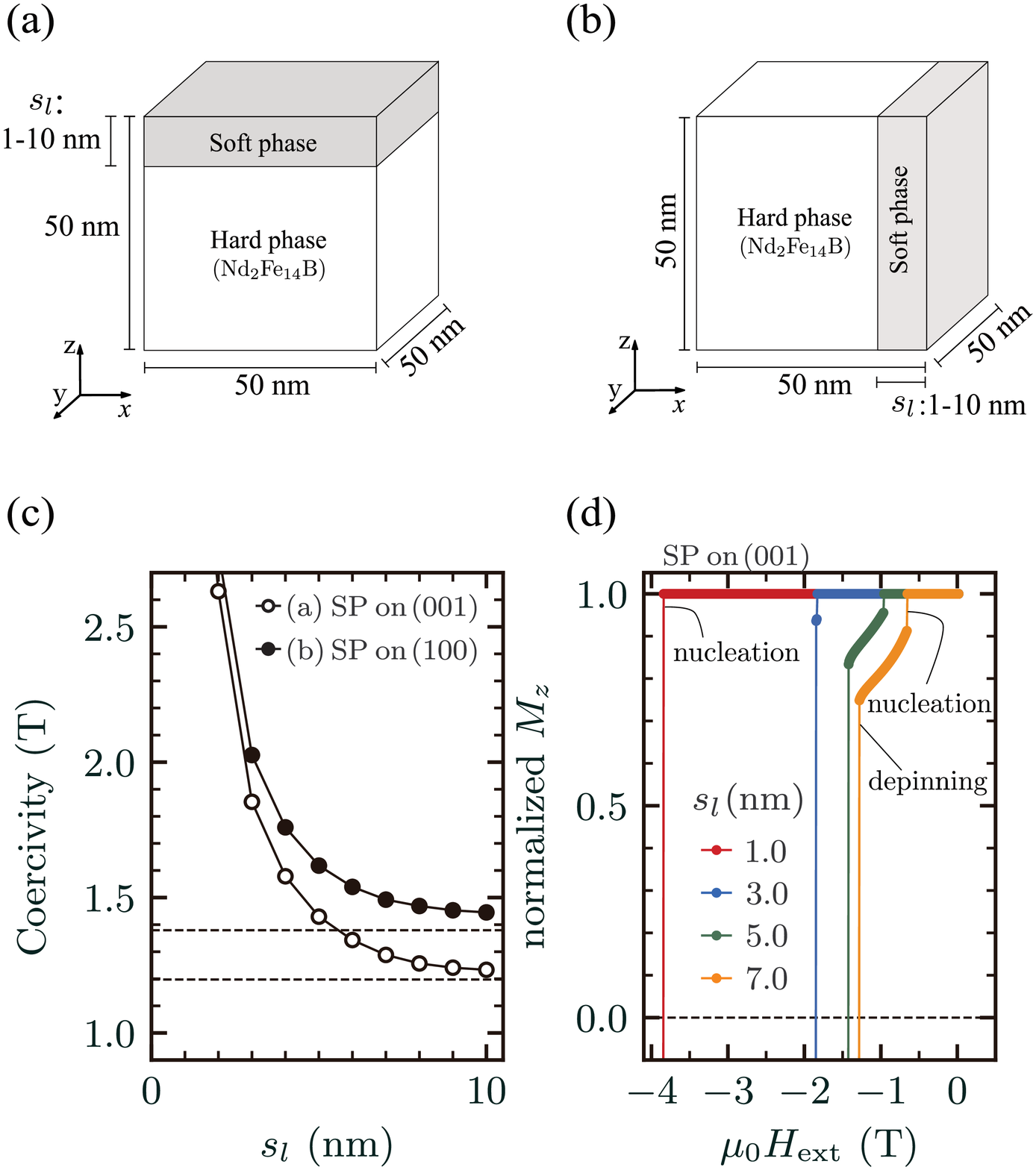}
\caption{\label{fig:hc}
Calculation models with open boundary conditions in which the soft magnetic phase (SP) is placed on (a) (001) surface and (b) (100) surface of the hard magnetic phase.
(c) Coercivity without dipole-dipole interaction evaluated using each model including the MC results at $400\,\rm K$ as a function of soft phase thickness, $s_l$.
Dashed lines denote the analytical results of the depinning type coercivity calculated form Eq.~\eqref{eq:hdwp}.
(d)
Hysteresis loops for the model (a) with four different $s_l$.
}
\end{figure}

Let us consider the effects of the anisotropy of $\mathcal{A}$ on the coercivity.
In actual rare-earth magnets which are composed of main rare-earth magnet phase and (magnetic or non-magnetic) grain boundary phase,
magnetization reversal is considered to occur by nucleation near the interface and by the DW propagation.
Thus, to study magnetization reversal in such a process, we carried out micromagnetic simulations for the two-phase models composed of the soft magnetic phase and the hard magnetic phase, depicted in Fig.~\ref{fig:hc} (a) and Fig.~\ref{fig:hc} (b).
The soft phase represents the grain boundary phase.
The two models (a) and (b) are the same if we do not take into account the anisotropy of $\mathcal{A}$ and the dipole-dipole interaction.


The simulations are based on the finite-difference method and the LLG equation\,\cite{nakatani_direct_1989, miltat_numerical_2007}:
\begin{eqnarray}
  \frac{d\bm{M}_i}{dt}=-\frac{|\gamma|}{1+\alpha^2}\left[\bm{M}_i\times \bm{H}_i^{\rm eff} 
  + \frac{\alpha}{|\bm{M}_i|} \bm{M}_i \times (\bm{M}_i\times \bm{H}_i^{\rm eff})  \right], \nonumber \\
\end{eqnarray}
where $\bm{M}_i$ is the magnetization vector of $i$th cell,
$\gamma$ is the gyromagnetic ratio constant, and $\alpha$ is Gilbert damping constant.
Both models (a) and (b) are discretized with a cubic cell of $(1.0\rm \, nm)^3$ and we set $|\gamma|=2.21\times 10^{5}\, \rm m/A \cdot sec$ (the value of free electron) and $\alpha=1$ (coercivity does not depend on $\alpha$).
Effective magnetic field on $i$th cell, $\bm{H}_i^{\rm eff}$, is defined as the derivative of the micromagnetic energy, $E^{\rm cont}_i$ (obtained from Eq.\eqref{eq:hami_cont}), with respect to $\bm{M}_i$\,\cite{tsukahara_effect_2018}:
\begin{eqnarray}
  \bm{H}_i^{\rm eff} &=& - \frac{1}{\mu_0} \frac{\partial E^{\rm cont}_i}{\partial \bm{M}_i} + H_{\rm ext} \bm{e}_u \nonumber \\
  &=& \sum_{j\in \rm n.n.} \frac{2\mathcal{A}_{ij}}{|\bm{M}_i|} 
  \frac{\bm{m}_{j}-\bm{m}_{i} }{ d_{ij}^2} \nonumber 
  +\frac{ 2K_1^i }{|\bm{M}_i|}(\bm{m}_i\cdot\bm{e}_u) \bm{e}_u+ H_{\rm ext} \bm{e}_u,\\
  \label{eq:heff}
\end{eqnarray}
where, $\mu_0$ is the magnetic permeability of the vacuum, $H_{\rm ext}$ is an external magnetic field,
$j$ represents six nearest neighbor cells of the $i$th cell, $\bm{m}_{i}=\bm{M}_i/|\bm{M}_i|$,
$\bm{e}_u$ is the unit vector of the easy axis,
$d_{ij}$ is the distance between the centers of $i$th and $j$th cells (i.e. $d_{ij}=1.0\rm \,nm$),
$\mathcal{A}_{ij}$ is the exchange stiffness costant,
and $K_1^i$ is the magnetic anisotropy constant (terms of $K_2^i$ and $K_4^i$ were omitted for simplicity).

In the present study, we determine the input model parameters in Eq.\eqref{eq:heff} from the MC results at $400\rm\,K$.
The model parameters in the hard phases are set to $|M_i|=1.38\,\rm T$, $K_1^i=2.63\,\rm MJ/m^3$,
$\mathcal{A}_{ij}= 12.21 \rm\, pJ/m$ for the pairs of $i_z=j_z$,
and $\mathcal{A}_{ij}= 9.10 \rm\, pJ/m$ for the pairs of $i_z\neq j_z$,
where $i_z(j_z)$ is the position of the $i(j)$th cell in the $z$-axis.
In the soft phases, the model parameters are set to $|M_i|=1.38\,\rm T$, $K_1^i=0\,\rm MJ/m^3$, and $\mathcal{A}_{ij}= 9.10 \rm\, pJ/m$ for all the pairs of $(i,j)$.
The difference of $\mathcal{A}_{ij}$ in the direction for the hard phases reflects the anisotropy of $\mathcal{A}$ in the MC results.
In addition, we assume $\mathcal{A}_{ij}= 9.10 \rm\, pJ/m$ for the bonds connecting the soft/hard interfaces.

By applying the fourth-order Runge-Kutta algorithm\,\cite{press_numerical_2007} to the LLG equation with the above-constructed models,
we simulated the magnetization reversal dynamics and then evaluated the coercivity.
In the simulation, we set the Runge-Kutta time step to $0.1\rm\, ps$, the magnetic field ($H_{\rm ext}$) is reduced by $25\rm\, mT$ at each field step, and the convergence condition under each field is when the average of magnetization torque, $|\bm{m}_i\times \bm{H}_i^{\rm eff}|$, is lower than $ 1.0\rm\, mOe$.
Also, to avoid the case of $\bm{m}_i \times \bm{H}_i^{\rm eff} \sim 0$, at the beginning of each field step, a disturbance is added to the magnetization vector of every cell as  $\bm{m}_i \rightarrow (\bm{m}_i + \bm{v})/| \bm{m}_i + \bm{v}|$, where $\bm{v}$ is a random vector of length $10^{-4}$.	
Using the simulation conditions, the magnetization reversal of the hard phase (not including the soft phase) occurs at $4.80\rm \,T$ which is consistent with the Stoner-Wohlfarth limit, $2K_1/M_s=4.789\rm \,T$.

In such the two-phase models, a magnetization reversal is expected to start from the soft phase.
Thus, we also examine the influence of soft phase thickness, $s_l$.
In Fig.~\ref{fig:hc}\,(c), we plot $s_l$ dependence of the coercivity for the models (a) and (b).
It is clearly seen that the coercivity of the model (a) is weaker than that of the model (b) regardless of $s_l$.
The relation of the coercivities between the two models is also consistent with that of the magsnitude of DW energy in each direction (see Fig.~\ref{fig:dfene}).
Since the models (a) and (b) are equivalent in the absence of anisotropy of $\mathcal{A}_{ij}$ in the hard phase, we can conclude that the difference of the coercivityu is attributed to the anisotropic $\mathcal{A}$.

It is also seen that as $s_l$ increases, the coercivity decreases and gradually approaches a certain value for each model, and the difference between two models increases.
As we will see in the following two paragraphs, this behavior is explained as a change in magnetization reversal mechanism from nucleation type to depinning type.
Here we define the nucleation type as a magnetization reversal of the whole system occurs by a nucleation which starts from nucleation at the surface of the soft phase without depinning at the interface of the soft and hard parts,
while in the depinning type, the reversed magnetization in the soft phase is pinned at the interface until the magnetic field reaches the threshold of the depinning.

Figure~\ref{fig:hc}\,(d) shows the hysteresis loops (showing the only upper part in the figure)
for the model (a) with four different $s_l$.
When the soft phase is thin ($s_l=1, 3\,\rm nm$), the magnetization of the hard phase is reversed at the same time as the nucleation at the soft phase (i.e., the nucleation type),
whereas when the soft phase is thick ($s_l=5, 7\,\rm nm$), coercivity is determined not from the nucleation but from the depinning (i.e., the depinning type).
The change between the nucleation type and depinning type were also pointed out in the previous studies for a one-dimensional model in which a soft phase of finite width is sandwiched between hard phases\,\cite{paul_general_1982,sakuma_micromagnetic_1990, kronmuller_micromagnetic_2002, mohakud_temperature_2016} and a corner defect model\,\cite{bance_influence_2014}.

In the one-dimentional model an analytical solution of the coercivity in the limit $s_l\rightarrow \infty$, which means the depinning type, was also proposed as follows\,\cite{sakuma_micromagnetic_1990, kronmuller_micromagnetic_2002}:
\begin{eqnarray}
H_{\rm dwp}  =
\frac{2K_1^{\rm H}}{|\bm{M}^{\rm H}|}
\frac{ 1- \frac{ \mathcal{A}^{\rm S}K_1^{\rm S} }{ \mathcal{A}^{\rm H}K_1^{\rm H}} }
   { \left( 1+
   \sqrt{ 
   \frac{ \mathcal{A}^{\rm S}|\bm{M}^{\rm S}|}{\mathcal{A}^{\rm H}|\bm{M}^{\rm H}|}
   } \right)^2 },
   \label{eq:hdwp}
\end{eqnarray}
where $|\bm{M}^{\rm S(H)}|$, $K_1^{\rm S(H)}$, and $\mathcal{A}^{\rm S(H)}$ are the model parameters of Eq.~\eqref{eq:heff} in soft (hard) phase.
%
%
To apply Eq.~\eqref{eq:hdwp} to the three-dimensional models, as the value of $\mathcal{A}^{\rm H}$, we use the value of $\mathcal{A}_{ij}$ in the direction perpendicular to the soft/hard interface, and calculate $H_{\rm dwp}$.
Dashed lines in Fig.~\ref{fig:hc}\,(c) indicate $H_{\rm dwp}$ for the models, which seems to explain well the lower limit of depinning type coercivity for our three-dimensional models.
Therefore, it is understood that the anisotropy of $\mathcal{A}$ has a large effect on the coercivity of the depinning type compared with that of the nucleation type.

\begin{figure}[t]
\vspace{5mm}
\includegraphics[width=8.6cm]{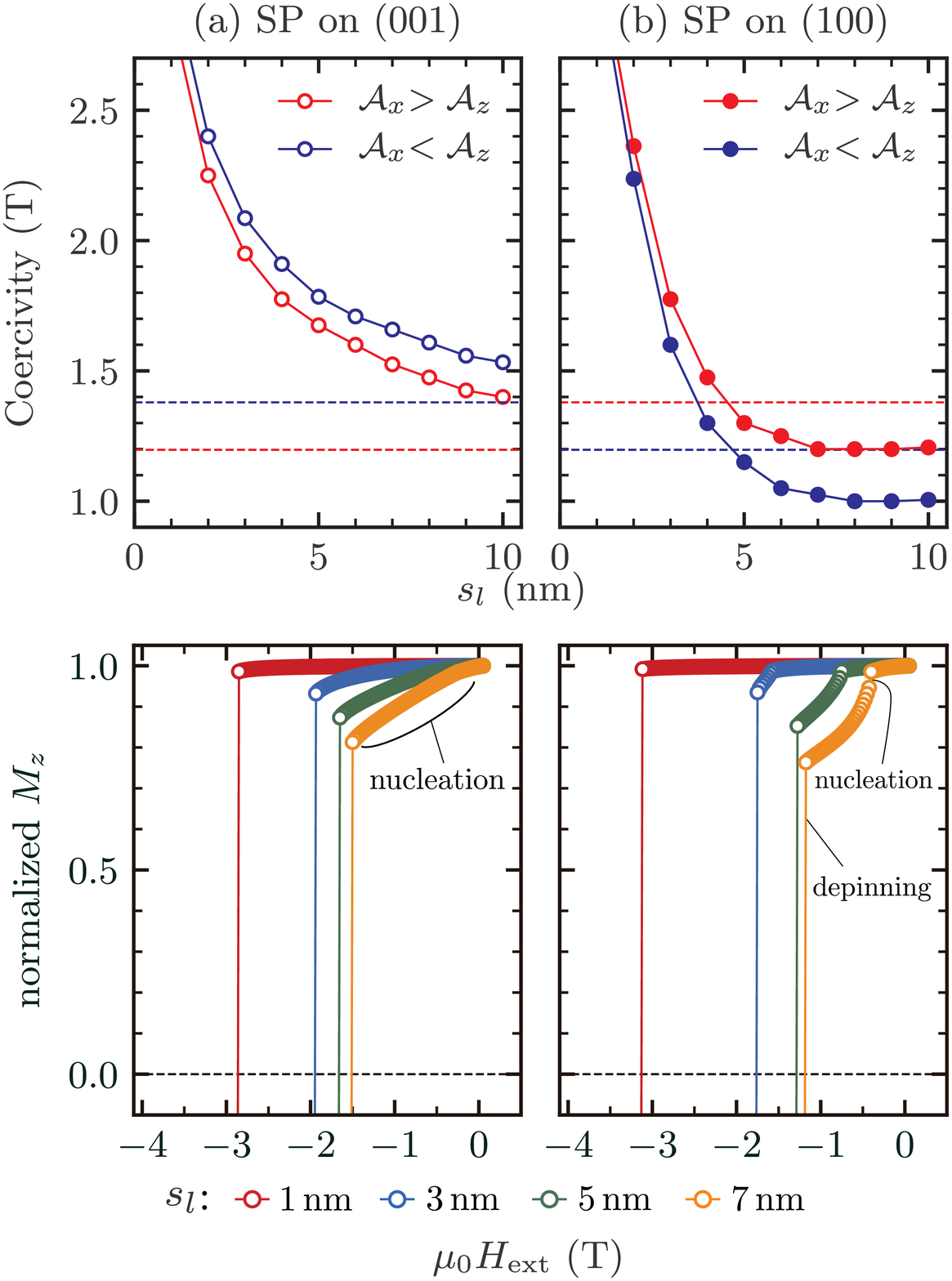}
\caption{\label{fig:hcdip}
(upper figures)
Coercivity with dipole-dipole interaction for the models (a) and (b) in Fig.~\ref{fig:hc}, as a function of soft phase thickness, $s_l$.
Dashed lines represent $H_{\rm dwp}$ calculated from Eq.~\eqref{eq:hdwp}.
(lower figures)
Hysteresis loops for the two models in the cases of $\mathcal{A}_x > \mathcal{A}_z $ (red circles in the upper figures) with four different $s_l$.
}
\end{figure}

Finally, we study the influence of the dipole-dipole interaction on the coercivity in the two models.
Magnetic field due to the dipole-dipole interaction is incorporated in $\bm{H}_i^{\rm eff}$ as the following form:
\begin{eqnarray}
\bm{H}_i^{\rm dip}= \sum_{j} \mathcal{K}(\bm{r}_{ij}) \bm{M}_j,
\end{eqnarray}
where $\mathcal{K}(\bm{r}_{ij})$ is the demagnetization tensor\,\cite{nakatani_direct_1989},
$\bm{r}_{ij}$ is the distance vector between $i$ and $j$ cells.
Here the strength of the dipole-dipole interaction is determined automatically according to the distance of the pair and the magnetization vector of the cells.
We calculate $\bm{H}_i^{\rm dip}$ of all the cells in $O(N_c{\rm log} N_c)$ computational time ($N_c$ is the total number of cells) by solving convolution integral using the fast Fourier transform method\,\cite{hayashi_calculation_1996}.
In the present study, we set the models (a) and (b) in Fig.~\ref{fig:hc} are cubic regardless of $s_l$ and 
$M^H=M^S$, and thus
the models (a) and (b) have the same shape magnetic anisotropy.

Upper figures of Fig.~\ref{fig:hcdip} show $s_l$ dependence of coercivity for the two models (a) and (b) with dipole-dipole interaction.
The red circles and the red dotted lines ($H_{\rm dwp})$ represent the values of the coercivity which are calculated using the same input parameters ($|M_i|$, $K_1^i$, and $\mathcal{A}_{ij}$) as those in Fig.~\ref{fig:hc} (c).
Here, the difference of coercivity in the models (a) and (b) in Fig.~\ref{fig:hcdip} is mainly attributed to the dipole-dipole interaction, and the anisotropic $\mathcal{A}$ is a secondary effect.
Because dipole-dipole interaction prefers to construct the DW along $z$-axis (type I),
in the model (b), the coercivity decreases compared with that in Fig.~\ref{fig:hc} (c).
Conversely in the model (a), the coercivity increases in the region of $s_l\geq 3 {\rm nm}$.
The dipole-dipole interaction inhibits to construct the DW (nucleation) in $xy$-plane.
For this reason, the coercivity with dipole-dipole interaction depends on the arrangement of the soft phase, which works contrary to the effect of anisotropic exchange stiffness, $\mathcal{A}_x > \mathcal{A}_z$.

These behaviors are confirmed from the hysteresis loops in lower figures in Fig.~\ref{fig:hcdip}.
In the model (b), magnetization reversal is clearly separated into the two parts, i.e., the small jump at the lower magnetic field where only the soft phase is reversed  (nucleation), and the jump at the higher magnetic field where the DW is depinned at the interface (depinning).
In contrast, in the model (a) they are not clearly distinguished.
That is, in the model (a), the magnetization reversal mechanism becomes to approach the nucleation type from the depinning type by the dipole-dipole interaction.
The dependence of the coercivity on the arrangement of the soft phase were similarly discussed in the most recent study (not including the anisotropy of $\mathcal{A}$)\,\cite{li_effects_2018}.

It is difficult to clarify the effect of anisotropic $\mathcal{A}$ on the coercivity under the dipole-dipole interaction by a simple comparison the models (a) and (b).
Thus, we exchanged the values of $\mathcal{A}_{x,y}$ and $\mathcal{A}_{z}$ in the hard phase.
Namely, we set the input parameters of the hard phase as $\mathcal{A}_{ij}= 9.10 \rm\, pJ/m$ for $i_z=j_z$,
and $\mathcal{A}_{ij}= 12.21 \rm\, pJ/m$ for $i_z\neq j_z$.
The values of coercivity under these conditions are plotted by the blue circles and the blue lines ($H_{\rm dwp}$) in the upper figures of Fig.~\ref{fig:hcdip}.
The anisotropy of $\mathcal{A}$ has a similar effect on the coercivity as the case without dipole-dipole interaction.
However, in the model (a), the difference in coercivity is relatively small.
The anisotropic $\mathcal{A}$ strongly affects the coercivity of the depinning type compared with the nucleation type.
On the other hand, The magnetization reversal in the model (a) is the nucleation type rather than the depinning type.
Therefore, we may conclude the dipole-dipole interaction in the model (a) suppresses the effect of the anisotropy of $\mathcal{A}$.



\section{Conclusion}

Regarding the exchange stiffness constant, $\mathcal{A}$, of Nd$_2$Fe$_{14}$B,
we examined the temperature and orientation dependences using the Monte Carlo simulations with the atomistic spin model constructed from the ab-initio calculation.
We also conducted the coercivity calculations based on the micromagnetics (LLG) simulations using the continuum model with the parameters obtained by the atomic scale MC results at finite temperatures.
In this way, we confirmed that the lattice structure in the atomic scale affects the coercivity as macroscopic physics.
We found that $\mathcal{A}(T)$ depends on the orientation of the crystal with respect to not only the amplitude but also the exponent $n_{x(z)}$ in the scaling behavior: $\mathcal{A}_{x(z)}(T)\propto M(T)^{n_{x(z)}}$; namely $\mathcal{A}_{x}(T) > \mathcal{A}_{z}(T)$ and $n_x < n_z$.
It is quantitatively confirmed that the anisotropic properties of $\mathcal{A}$ come from the weak exchange couplings between Nd and Fe atoms and the layered structure of Nd atoms.
Moreover, we also found that the anisotropic $\mathcal{A}(T)$ affects the coercivity of the depinning mechanism.

We focused on only Nd$_2$Fe$_{14}$B magnet in the present paper.
However, the essence of anisotropic $\mathcal{A}$ comes from the weak exchange coupling between rare-earth atoms and transition metals and the layered structure of rare-earth atoms.
Thus, the features discussion for Nd$_2$Fe$_{14}$B are probably applicable to other rare-earth magnets.
In fact, it was pointed out by ab-initio calculations that $\mathcal{A}$ have strong anisotropy for YCo$_5$\,\cite{belashchenko_anisotropy_2004} and Sm(Fe,Co)$_{12}$\,\cite{fukazawa_first-principles_2018}.

Let us consider the coercivity from the viewpoint of the exchange spring magnet\,\cite{kneller_exchange-spring_1991,skomski_giant_1993} which is composed of hard and soft phase and expected to realize the highest performance magnet.
Because realization of high performance requires a large coercivity and a large thickness of soft phase,
the model (a) with $\mathcal{A}_x<\mathcal{A}_z$ (Fig.~\ref{fig:hcdip} (a) blue line) is the most suitable conditions in our modelings.
Most strong permanent magnets, Nd$_2$Fe$_{14}$B, YCo$_5$ and also $L1_0$-type magnet (CoPt, FePd, FePt) cannot reproduce the same condition because $\mathcal{A}_x > \mathcal{A}_z$\,\cite{belashchenko_anisotropy_2004},
whereas Sm(Fe,Co)$_{12}$ would do because the anisotropy as $\mathcal{A}_x < \mathcal{A}_z$\,\cite{fukazawa_first-principles_2018}.
Therefore, Sm(Fe,Co)$_{12}$ and other $R$(Fe,Co)$_{12}$-type compounds ($R$ is a rare-earth element) may have higher potential to realize strong performance exchange spring magnet rather than the other magnets.

Finally, we point out another source of the anisotropy.
In a recent experiment, it has been observed that
the grain boundary phase takes different crystal structures and chemical compositions depending on the orientation with the Nd$_2$Fe$_{14}$B main phase,
i.e., the Nd-rich crystalline paramagnetic phase form on the $xy$-plane of the main phase,
whereas the Fe-rich amorphous ferromagnetic phase in the plane parallel to the $z$-axis\,\cite{sasaki_structure_2016}.
%
In the present paper, we have studied the effect of anisotropy of $\mathcal{A}$ on coercivity and of the orientation of the interface with the soft phase changes by using the same interaction for the interface.
However, if the chemical structure is different, the exchange interaction would be a difference due to another source of the anisotropy,
which is studied with information of the structure in the future.

In the continuum model, Eq.~\eqref{eq:hami_cont}, we used 
the values of $K_u$ and $\mathcal{A}$ obtained by the MC simulations which are the values of the bulk system. There,
changes in atomic scale of magnetic anisotropy\,\cite{moriya_first_2009,toga_effects_2015,tatetsu_first-principles_2016} and exchange coupling\,\cite{sabiryanov_magnetic_1998,toga_first_2011,ogawa_negative_2015, umetsu_first-principles_2016} near the interface or surface were not taken into consideration.
The influences of interface and surface are important for the coercivity\,\cite{hirosawa_perspectives_2017}.
Thus, the accuracy multiscale analysis needs further development of connecting scheme from atomistic spin model to macrospin model would be necessary.

As another point noted is the range of exchange interaction.
In the atomistic spin model, Eq.~\eqref{eq:hami}, of the present study, we omitted the long-range contribution of $\tilde{J}_{ij}$ for simplicity and reduction of calculation cost.
However, the recent study\,\cite{da_silva_junior_domain_2017} reported that RKKY-type exchange coupling significantly effects on the DW width, and pointed out the importance of the long-range contribution.
We also found that, by incorporating long-range exchange couplings up to $10.6\,\rm\AA$, the difference between type I and type II of $E_{\rm dw}$ at $400\,\rm K$ reduces from $11.3\,\%$ (in Fig.~\ref{fig:dfene}) to $4.4\,\%$.
However, to make study the effect clearly, we need precise information of the interaction at the long distance, and we postpone to study this problem later.


Although there are still problems that must be concerned,
we believe that the present paper will be helpful to elucidate coercivity mechanism in rare-earth permanent magnets.
%

\begin{acknowledgments}
We acknowledge collaboration and fruitful discussions with Taro Fukazawa, Taichi Hinokihara, Shotaro Doi, Munehisa Matsumoto, Hisazumi Akai, and Satoshi Hirosawa.
This work was partly supported by Elements Strategy Initiative Center for Magnetic Materials (ESICMM) under the auspices of MEXT; by MEXT as a social and scientific priority issue (Creation of New Functional Devices and High-Performance Materials to Support Next-Generation Industries; CDMSI) to be tackled by using a post-K computer.
The computation was performed on Numerical Materials Simulator at NIMS; the facilities of the Supercomputer Center, the Institute for Solid State Physics, the University of Tokyo; the supercomputer of ACCMS, Kyoto University.
\end{acknowledgments}

\bibliographystyle{apsrev4-1} 
\bibliography{07.stiffness} 

%


\end{document}